\documentclass{article}

\usepackage[final]{neurips_2025}

\usepackage[utf8]{inputenc} 
\usepackage[T1]{fontenc}    
\usepackage{hyperref}       
\usepackage{url}            
\usepackage{booktabs}       
\usepackage{amsfonts}       
\usepackage{nicefrac}       
\usepackage{microtype}      
\usepackage{xcolor}         
\usepackage{graphicx}
\usepackage[linesnumbered,ruled,vlined]{algorithm2e}
\usepackage{amsmath}
\usepackage{algorithmic}
\usepackage{subcaption}

\title{FlowHFT: Imitation Learning via Flow Matching Policy for Optimal High-Frequency Trading under Diverse Market Conditions}

\author{
\hspace{-1cm}
Yang Li$^{1}$, \, 
Zhi Chen$^{1}$,\,
Steve Yang$^{1,\dagger}$\\
\hspace{-1cm}
$^{1}$Stevens Institute of Technology\quad
\\
\hspace{-1cm}
\texttt{\{yli269, zchen100, syang14\}@stevens.edu}
}

\begin{document}

\maketitle

\begin{abstract}
High-frequency trading (HFT) is an investment strategy that continuously monitors market conditions and places bid and ask orders at nanosecond speeds. Traditional HFT approaches fit models using historical data and assume future market states will follow similar patterns, limiting the effectiveness of any single model to the specific conditions under which it was trained. Additionally, such models achieve optimal performance only under restrictive assumptions, including specific stochastic processes for asset prices, stable order flows, and the absence of sudden volatility spikes. However, real-world financial markets are dynamic, heterogeneous, and frequently volatile.
To address these challenges, we propose \textbf{FlowHFT}, a novel \textbf{imitation learning framework} based on a \textbf{flow matching policy}. FlowHFT simultaneously learns strategies from numerous expert models, each of which is proficient in specific market scenarios. As a result, our framework adaptively adjusts investment decisions according to prevailing market conditions. Furthermore, FlowHFT incorporates a grid-search fine-tuning mechanism, enabling it to refine strategies and achieve superior performance even under complex or extreme market conditions where expert strategies may become suboptimal.
We evaluate FlowHFT across various market environments. Our study is the first to demonstrate the successful application of flow matching policy within stochastic market environments. 
Our findings show that FlowHFT can simultaneously learn trading strategies for each market condition in a single model. Notably, our unified framework consistently outperforms the best-performing expert strategy in each tested market scenario.
\end{abstract}

\section{Introduction}
Market making (MM) is a trading strategy where participants place both buy and sell orders to profit from difference in the bid-ask spread. By providing liquidity through these actions, market makers also enhance market efficiency ~\cite{madhavan2000market, harris2002trading}.
High-frequency trading (HFT) is one type of market making that executes orders at high speeds, often measured in milliseconds or microseconds. This rapid execution allows traders to earn profit on very small price changes within short timeframes
~\cite{avellaneda2008high,cartea2014buy,ait2017high}.

Traditional HFT strategies, such as the Avellaneda-Stoikov (AS) model ~\cite{avellaneda2008high} and the Guéant-Lehalle-Fernandez-Tapia (GLFT) model
~\cite{gueant2013dealing}, rely on historical market data to calibrate their parameters. These calibrated models are used to place orders in subsequent market. However, the effectiveness of traditional HFT models is constrained by two key limitations. If future market conditions diverge substantially from the historical period used for calibration, the optimized parameters may become ineffective and lead to suboptimal performance. 
Moreover, these models are often built on optimization frameworks requiring strict assumptions about market behavior (e.g., specific stock price processes, stable order flow, no sudden jumps) that reflect idealized rather than real-world scenarios. Consequently, when actual market conditions violate these assumptions—a common occurrence—these traditional models struggle to adapt and tend to produce suboptimal trading outcomes.


In the recent literature, Reinforcement learning (RL) has been applied to high-frequency trading, framing the problem as an agent (the trading algorithm) interacting with an environment (the market). The agent observes the market state, takes actions and receives rewards based on metrics like profit and loss. Through this interaction, the RL aims to learn a policy—a strategy—that maximizes cumulative rewards. 
However, RL often focuses on optimizing only single-step actions, which can lead to compounding error ~\cite{ganesh2018deep,kober2013reinforcement,zheng2024reinforcement} for instance, slightly suboptimal actions at each step might accumulate, resulting in a large unwanted position.
Effective HFT necessitates a long-term view, because profitability typically relies on accumulating numerous small gains from individual steps. 
Moreover, each action taken interacts with the market environment, thereby impacting the feasibility and profitability of subsequent moves.

To address the challenges associated with traditional HFT methods, we propose FlowHFT, a novel imitation learning framework based on flow matching policy ~\cite{chi2024diffusionpolicy,sohl2015deep,ho2020denoising,song2020score,janner2022planning,wang2022diffusion}. This framework consists of two primary components. The first component utilizes imitation learning to develop a pre-trained model for observing the market state and generate corresponding trading actions. We simulate diverse market scenarios and evaluate various traditional models within each. For every scenario, the model achieving the best performance is designated as the "expert". FlowHFT then learns by imitating the actions of these designated experts across all scenarios, effectively integrating the strongest known strategy for each market condition into one single, comprehensive model. 
Imitation learning is useful because these models can in fact generate optimal solutions in certain scenarios. Even in those scenarios where they only generate suboptimal solutions, FlowHFT can still learn their strategies and improve on them subsequently. 
The second component uses a grid search mechanism to fine-tune the action proposed by the pre-trained model. 
The initial action proposed by the pre-trained model, refined by the adjustment from the fine-tuning stage, constitutes the final action.

To the best of our knowledge, this paper presents the first flow matching policy based imitation learning approach to a financial stochastic control problem. The results show that the FlowHFT framework integrates the knowledge of multiple expert strategies into a single adaptive model. When market scenarios change, the framework demonstrates adaptability by generating actions appropriate to the prevailing conditions, effectively leveraging patterns learned from relevant expert demonstrations. 
Additionally, FlowHFT is designed for rapid inference, enabling the generation of trading actions at millisecond speeds for HFT tasks.
Furthermore, the grid search mechanism refines these generated actions. The experimental results indicate that FlowHFT consistently achieves superior performance compared to the best results obtained from the baseline models individually optimized for each specific market scenario.

The results indicate that our model exhibits significant robustness to market conditions involving sudden price jumps. We attribute this resilience primarily to our method's output: it generates sequences of actions over a planning horizon, rather than isolated single-step decisions. By producing action sequences, the model inherently considers a near-term trajectory and allows for smoother adjustments, effectively taking into account the broader consequences of initial moves. This sequential planning perspective helps mitigate the compounding error, aligning actions more effectively over time and thus enhancing strategic stability and performance during volatile events like jumps.

\begin{figure}[htp]
\caption{Visualization of FlowHFT's Input Observation Sequence and Output Action Sequence}
\label{fig:seeandaction}
\centering
\includegraphics[width=0.75\textwidth]{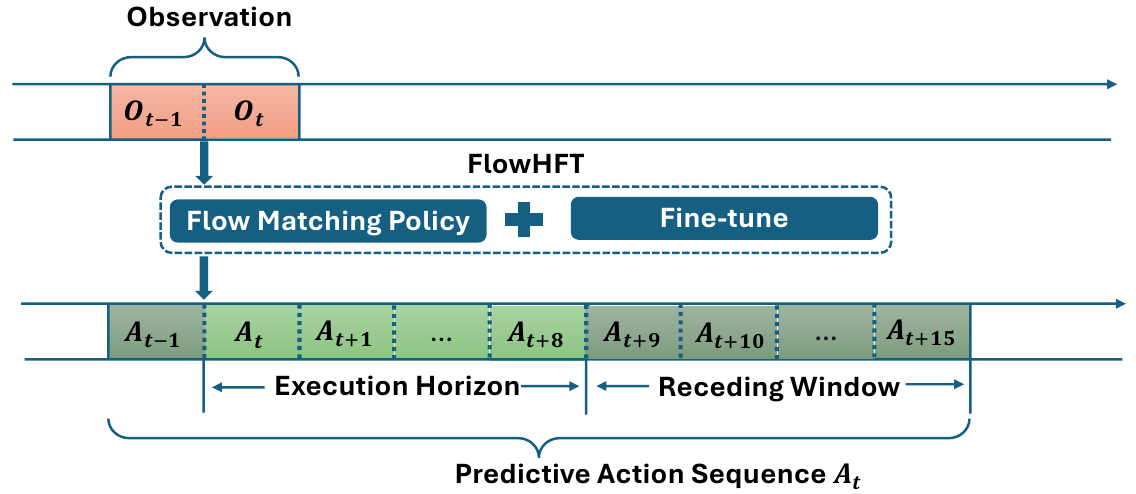}
\end{figure}

\section{Preliminaries}
\label{sec:Preliminaries}
In this section, we formulate the high frequency trading task mathematically.
High-frequency trading utilizes automated algorithms for order execution at extremely high speeds, often operating within millisecond or nanosecond timescales. The objective is to continuously place buy (bid) and sell (ask) limit orders for a specific financial instrument, while aiming to profit from the bid-ask spread.

We can model the HFT market making task as a stochastic control process \cite{avellaneda2008high, gueant2013dealing} defined over a set of discrete time steps $\mathbb{T}=\{0, 1, \dots, T\}$. The \textbf{observation state space} $\mathcal{O}$ captures relevant information at time $t \in \mathbb{T}$. An observation state $O_t \in \mathcal{O}$ typically includes observable market information $L_t$ (often derived from the Limit Order Book, LOB, including stock prices and bid-ask spread) and the agent's information $Z_t$, including balance, current inventory level , along with the time $t$ itself, such that $\mathcal{O} \subseteq \mathcal{L} \times \mathbb{Z} \times \mathbb{T}$, where $\mathcal{L}$, $\mathbb{Z}$ is the space of market information and agent information, respectively. We generally assume the state $S_t$ satisfies the Markov property.

Within this framework, the market maker chooses an \textbf{action} $A_t$ from the \textbf{action space} $\mathcal{A}$. A common action involves setting the agent's bid and ask quotes, often parameterized by spreads $(\delta^b_t, \delta^a_t)$ relative to a reference price $p^{ref}_t$ derived from the market state $L_t$. The system's evolution is governed by stochastic transition probabilities $P(O_{t+1} | O_t, A_t)$, defining the likelihood of moving to the next observation state $O_{t+1}=(L_{t+1}, I_{t+1}, t+1)$ given the current state and action. This transition reflects changes in both market conditions $L_{t+1}$ and the agent's inventory $I_{t+1}$ due to order executions and market activity. The agent seeks an optimal \textbf{policy} $\pi: \mathcal{O} \to \mathcal{A}$, a rule mapping states to actions ($A_t = \pi(O_t)$), that maximizes a defined \textbf{objective function}, $J(\pi)$. A standard objective is to maximize the expected final value, often combining terminal cash wealth $W_T$ with a penalty $\phi(I_T)$ for inventory risk at the horizon $T$. The optimization problem is thus:
\begin{equation} \label{eq:mm_objective}
\max_{\pi} J(\pi) = \max_{\pi} \mathbb{E}^{\pi} \left[ W_T - \phi(I_T) \mid O_0 \right]
\end{equation}
Where, $\mathbb{E}^{\pi}[\cdot]$ denotes the expectation under policy $\pi$ starting from the initial observation state $O_0 = (L_0, I_0, 0)$. Solving for the optimal policy $\pi^* = \arg\max_{\pi} J(\pi)$ requires dynamically balancing profitability against inventory and adverse selection risks in a complex, stochastic environment.

\section{Architecture of FlowHFT}
\label{sec:Architecture of FlowHFT}

\subsection{Backbone of FlowHFT: Flow Matching Policy}
\label{sec:flow_matching_policy}

\begin{figure}[htp]
\caption{This figure illustrates the training of a flow matching policy. The inputs are various market situations. For each situation, the best expert is chosen from a pool of models. These optimal market-expert pairings are then used to train the flow matching policy, which operates as a vector field. This field maps the distribution of market conditions to a corresponding optimal action distribution.}
\label{}
\centering
\includegraphics[width=1\textwidth]{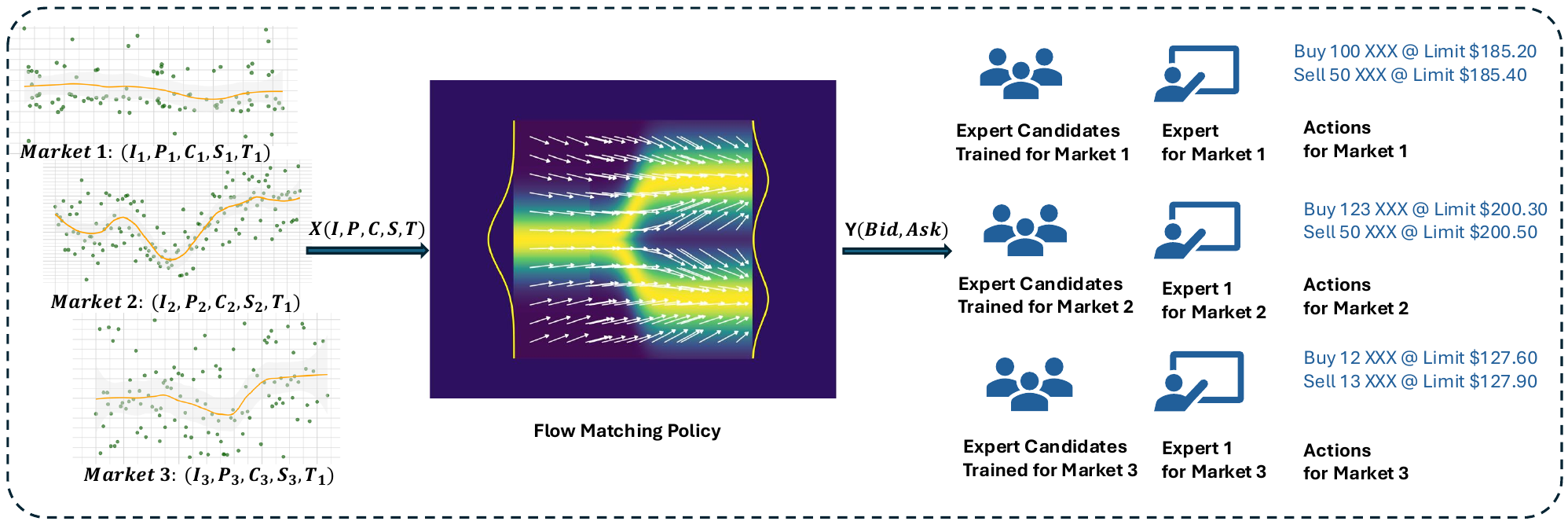}
\end{figure}

The core of FlowHFT is a conditional flow matching policy \cite{lipman2022flow}, denoted $\pi_{\theta}(A_{t+1}\mid O_t)$, which learns to map sequences of market observations $O_t$ to trading actions $A_{t+1}$. This policy is trained via imitation learning from expert demonstrations. We draw inspiration from recent successes of flow matching in domains like robotics, where these models effectively learn complex, continuous control policies from expert data by mapping high-dimensional inputs to sophisticated, sequential actions.

Our objective is to train a neural network $v_{\theta}(a, t \mid O_t)$ to parameterize a conditional vector field. This vector field defines a probability flow that transports samples from a simple prior distribution $p_{\mathrm{prior}}(a_0)$ (e.g., Gaussian noise) to the distribution of expert trading action sequences $p_{\mathrm{expert}}(a_E \mid O_t)$, conditioned on the observed market state $O_t$. Following seminal work on flow matching, we define a time-dependent probability path $p_t(a \mid a_0, a_E, O_t)$ that interpolates between an initial random action sequence $a_0$ and an expert action sequence $a_E$. A simple linear interpolation is given by:
$
a_t = (1 - t)\,a_0 + t\,a_E,
    \quad
    u_t(a_t \mid a_0, a_E) = a_E - a_0
$.
The training procedure (Algorithm~\ref{alg:fm_training}) minimizes the discrepancy between the network's predicted vector field $v_{\theta}(a_t, t \mid O_t)$ and the target vector field $u_t(a_t \mid a_0, a_E)$. Specifically, given a market state $O_t$, the model samples a random initial action sequence $a_0 \sim p_{\mathrm{prior}}$ and a corresponding expert action sequence $a_E \sim p_{\mathrm{expert}}(\cdot \mid O_t)$. It then samples a time $t \sim \mathcal{U}(0,1)$ and forms an interpolated action sequence $a_t$ using the defined path. The network is trained by minimizing the following Flow Matching (FM) loss:

\begin{equation}
\mathcal{L}_{\mathrm{FM}}(\theta) = \mathbb{E}_{t, O_t, a_0, a_E, a_t \sim p_t(a\mid a_0,a_E,O_t)} \left[ \| v_{\theta}(a_t, t \mid O_t) - u_t(a_t \mid a_0, a_E) \|^2 \right].
\label{eq:fm_loss}
\end{equation}

This objective effectively teaches the network $v_{\theta}$ to generate the flow that transforms noise into expert-like HFT action sequences conditioned on the market state.

\begin{equation}
    a_{k+1} = a_k + \frac{1}{N} \; v_{\theta}(a_k, t_k \mid O_t),
    \quad
    t_k = k/N,
    \; k = 0, \dots, N-1.
\label{eq:fm_inference}
\end{equation}

\begin{algorithm}[htbp]
\SetKwInput{KwData}{Input}
\SetKwInput{KwParam}{Parameters}
\SetKwRepeat{Do}{do}{until}
\caption{Flow Matching Policy Training }
\label{alg:fm_training}
\KwParam{Neural network $v_{\theta}(a, t | O_t)$ with parameters $\theta$, learning rate $\eta$}
\KwData{Expert demonstration data $D_{\text{expert}}$ containing pairs of $(O_t, a_E(O_t))$} 

\Do{not converged}{
    Sample a market observation $O_t$ and corresponding expert action sequence $a_E \sim p_{\text{expert}}(\cdot | O_t)$ from $D_{\text{expert}}$ (or generate via simulation/expert)\;
    Sample an initial action sequence $a_0 \sim p_{\text{prior}}(\cdot)$ (e.g., $a_0 \sim \mathcal{N}(0, I)$)\;
    Sample a time $t_{\text{sample}} \sim \mathcal{U}(0,1)$\;
    
    $a_{t_{\text{sample}}} \leftarrow (1 - t_{\text{sample}})a_0 + t_{\text{sample}}a_E$\;
    $u_{t_{\text{sample}}} \leftarrow a_E - a_0$\;
    
    $\hat{v}_{t_{\text{sample}}} \leftarrow v_{\theta}(a_{t_{\text{sample}}}, t_{\text{sample}} | O_t)$\;
    $\mathcal{L} \leftarrow \| \hat{v}_{t_{\text{sample}}} - u_{t_{\text{sample}}} \|^2$ 
    
    $\theta \leftarrow \theta - \eta \nabla_{\theta} \mathcal{L}$\;
}
\end{algorithm}

At inference (Algorithm~\ref{alg:fm_sampling}), to generate an action sequence conditioned on a live market observation $O_t$, the policy starts with an action sequence $a_0 \sim p_{\mathrm{prior}}$. It then iteratively refines this sequence by solving the ordinary differential equation:
$
\frac{da}{dt} = v_{\theta}(a, t \mid O_t),
$
from $t=0$ to $t=1$. This is typically achieved using a numerical ODE solver (e.g., Euler or Heun method) over a small number of discretization steps $N$. This process transforms the initial noise sequence into a coherent, market-aware sequence of high-frequency trading actions.

\begin{algorithm}[htbp]
\SetKwInput{KwData}{Input}
\SetKwInput{KwParam}{Parameters}
\caption{Flow Matching Policy Inference}
\label{alg:fm_sampling}
\KwParam{Trained policy network $v_{\theta}(a, t | O_t)$, Number of discretization steps $N$}
\KwData{Current market observation $O_t^{\text{current}}$}
$a \sim p_{\text{prior}}(\cdot)$ (e.g., $a \sim \mathcal{N}(0, I)$) \\
$t_{\text{current}} \leftarrow 0$\;
$\Delta t_{\text{step}} \leftarrow 1/N$\;

\For{$k \leftarrow 0$ \KwTo $N-1$}{
    $v_k \leftarrow v_{\theta}(a, t_{\text{current}} | O_t^{\text{current}})$  \\
    $a \leftarrow a + v_k \cdot \Delta t_{\text{step}}$\\
    $t_{\text{current}} \leftarrow t_{\text{current}} + \Delta t_{\text{step}}$\;
}

\KwOut{$a$} 
\end{algorithm}

\newpage

\subsection{Fine-tune Pretrained Model}
In the first component, a flow matching policy learns to imitate the strategies of various experts. So it is a general, pre-trained policy. 
The second component of our framework aims to fine-tune this pre-trained policy to surpass the performance of experts.
To enhance the framework's real-time adaptability to immediate market dynamics, we apply a linear transformation to the actions proposed by the pre-trained model. The parameters of this transformation are rapidly calibrated using the most current market information. This process is more computationally lightweight and efficient compared to recalibrating the traditional HFT models. The output of this calibrated linear transformation is then combined with the initial action from the pre-trained model to yield the final, refined trading decision.

Specifically, this fine-tuning uses a linear transformation of the form $(a \cdot \text{output} + b$) to the action sequence output by the pretrained model. Specifically, if the pretrained model outputs a sequence of planned actions $({a_{t+1}, \dots, a_{t+T_{\text{pred}}}})$ over the prediction horizon, where each action $(a_i)$ is a vector representing the bid and ask offsets for time $(t+i)$ (e.g., $(a_i = [\delta_{\text{bid}, i}, \delta_{\text{ask}, i}])$), the fine-tuned action sequence $({a'{t+1}, \dots, a'{t+T_{\text{pred}}}})$ is computed as $a'i = a \cdot a_i + b)$ for each step $(i)$ from 1 to $(T{\text{pred}})$. Here, $(a)$ is a scalar scaling factor applied element-wise to each vector $(a_i)$, and $(b)$ is an additive vector offset applied to each vector $(a_i)$. Both the scalar $(a)$ and the vector $(b)$ (with dimensions matching the action vector, e.g., 2 for bid/ask offsets) are calibrated using a validation set. This adjusted action sequence $(a')$ is then used by the policy (typically by executing the first action $(a'_{t+1})$) in the environment.

\begin{figure}[htp]
\caption{This figure shows the action generation. The neural network of flow matching policy (left) provides initial bid/ask outputs. A grid search (right) identifies adjustments ($\Delta$bid,$\Delta$ask) to refine these actions. Combine the initial result and adjustment together to get the final results.}
\label{}
\centering
\includegraphics[width=0.8\textwidth]{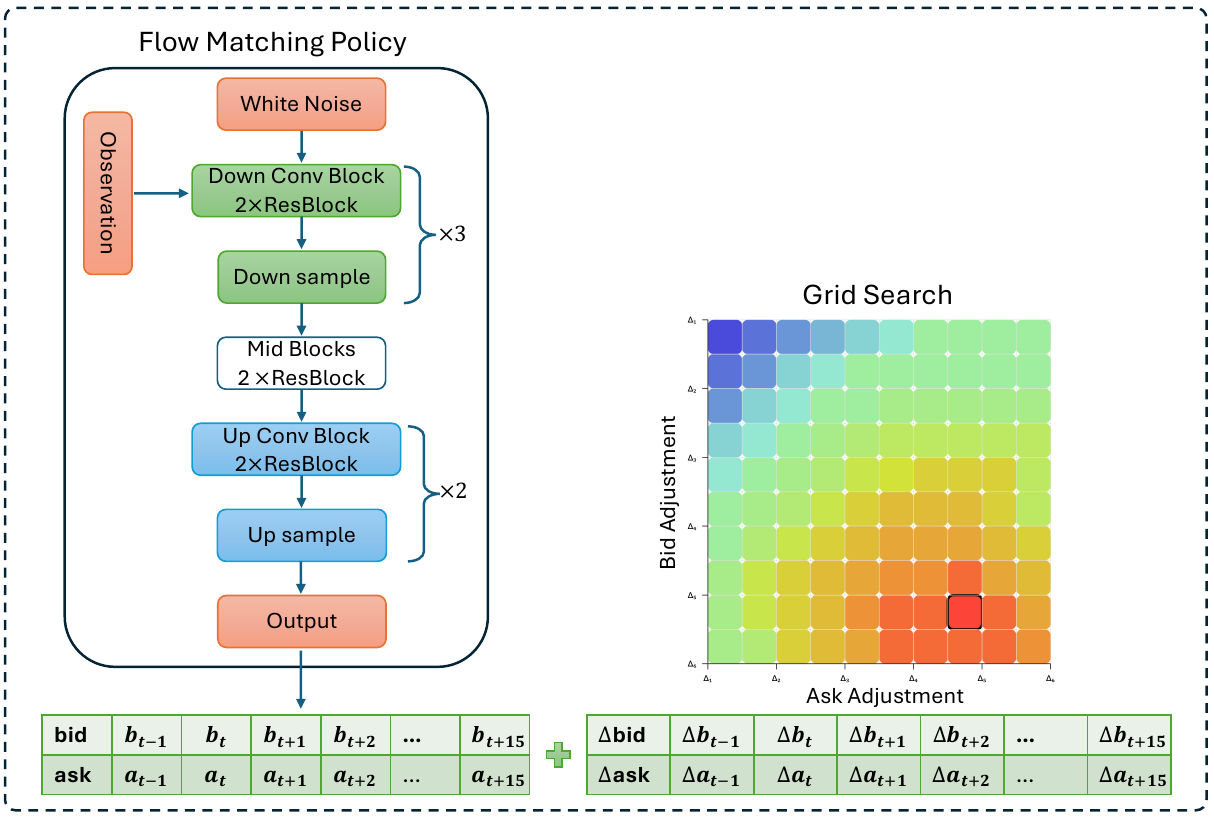}
\end{figure}

\section{Data of Imitation Learning}
\label{sec:data of Imitation Learning}
FlowHFT is an imitation learning framework. Therefore, it is crucial to prepare high-quality learning material.
In this section, we firstly detail how to generate a set of various market scenarios. Secondly, for each individual scenario, we evaluate a pool of candidate experts—including established traditional algorithms and reinforcement learning agents—and designate the one exhibiting the best performance as the "expert" for that context. We use the actions generated by the expert's strategy to create state-action pairs for training FlowHFT.

\subsection{Market Simulation}


We model the mid-price $S_t$ as a jump-diffusion process~\cite{merton1976option}, capturing both continuous price movements and sudden jumps. 
The stochastic differential equation (SDE) governing the mid-price is given by:

\begin{equation}
    \label{eq:mid_price}
dS_t = S_{t^-} \left( \mu dt + \sigma dB_H(t) \right) + S_{t^-} (e^{J} - 1) dN_t
\end{equation}

where $\mu$ represents the drift coefficient, signifying the expected return of the asset. The asset's volatility is denoted by $\sigma$, which quantifies the fluctuations in its price. The term $dB_H(t)$ corresponds to the increment of a fractional brownian motion process, capturing the continuous component of price movement. Furthermore, $J$ denotes the size of the jump, following a normal distribution $N(\mu_J, \sigma_J^2)$, which specifies the mean and variance of the jump. Lastly, $dN_t$ is a Poisson process with intensity $\lambda_J$, modeling the occurrence of jumps, thereby reflecting sudden shifts in the asset's price.

Next we discuss simulation of order arrival. 
Order arrivals are modeled using mutually exciting Hawkes processes~\cite{bacry2015hawkes,hawkes1971point}, effectively capturing the self-exciting and the cross-exciting nature of financial markets. 
"self-exciting" effects indicate that prior buy (or sell) orders increase subsequent arrivals of the same order type, respectively. "Cross-exciting" effects describe how buy orders influence sell order arrivals and vice-versa.

The intensity of buy and sell orders, $\lambda_b(t)$ and $\lambda_a(t)$ respectively, are defined as:
\begin{align}
\lambda_b(t) &= \mu_b + \sum_{t_i \in \mathcal{N}_b} \alpha_{bb} e^{-\beta (t - t_i)} + \sum_{t_j \in \mathcal{N}_a} \alpha_{ba} e^{-\beta (t - t_j)}, \label{eq:lambda_b} \\
\lambda_a(t) &= \mu_a + \sum_{t_i \in \mathcal{N}_a} \alpha_{aa} e^{-\beta (t - t_i)} + \sum_{t_j \in \mathcal{N}_b} \alpha_{ab} e^{-\beta (t - t_j)}, \label{eq:lambda_a}
\end{align}
where, $\mu_b$ and $\mu_a$ are the baseline intensities for buy and sell orders, respectively. The parameters $\alpha_{bb}$ and $\alpha_{aa}$ describe the self-exciting effects of buy and buy, and sell and sell orders, respectively, while $\alpha_{ba}$ and $\alpha_{ab}$ represent the cross-exciting effects between buy and sell orders. The parameter $\beta$ denotes the decay rate, which influences how quickly the effects of past events diminish over time. Finally, $N_b$ and $N_a$ are the sets containing the timestamps of past buy and sell order arrivals, which are used to model the timing of these events.

We create market scenarios with varying levels of liquidity—categorized as high, medium, and low. Furthermore, to test the model performance under stress, we introduce scenarios featuring sudden market changes and periods of significantly increased volatility, allowing us to examine how effectively each model responds to dynamic and challenging environments. We simulate different market environments by adjusting the following Simulation parameters. Combining these parameters yields $3 \times 3 \times 3 \times 2 \times 2 \times 3  \times 1 = 324$  parameter combinations.

\begin{figure}[htp]
\caption{Parameters settings of market simulation}
\label{}
\centering
\includegraphics[width=0.82\textwidth]{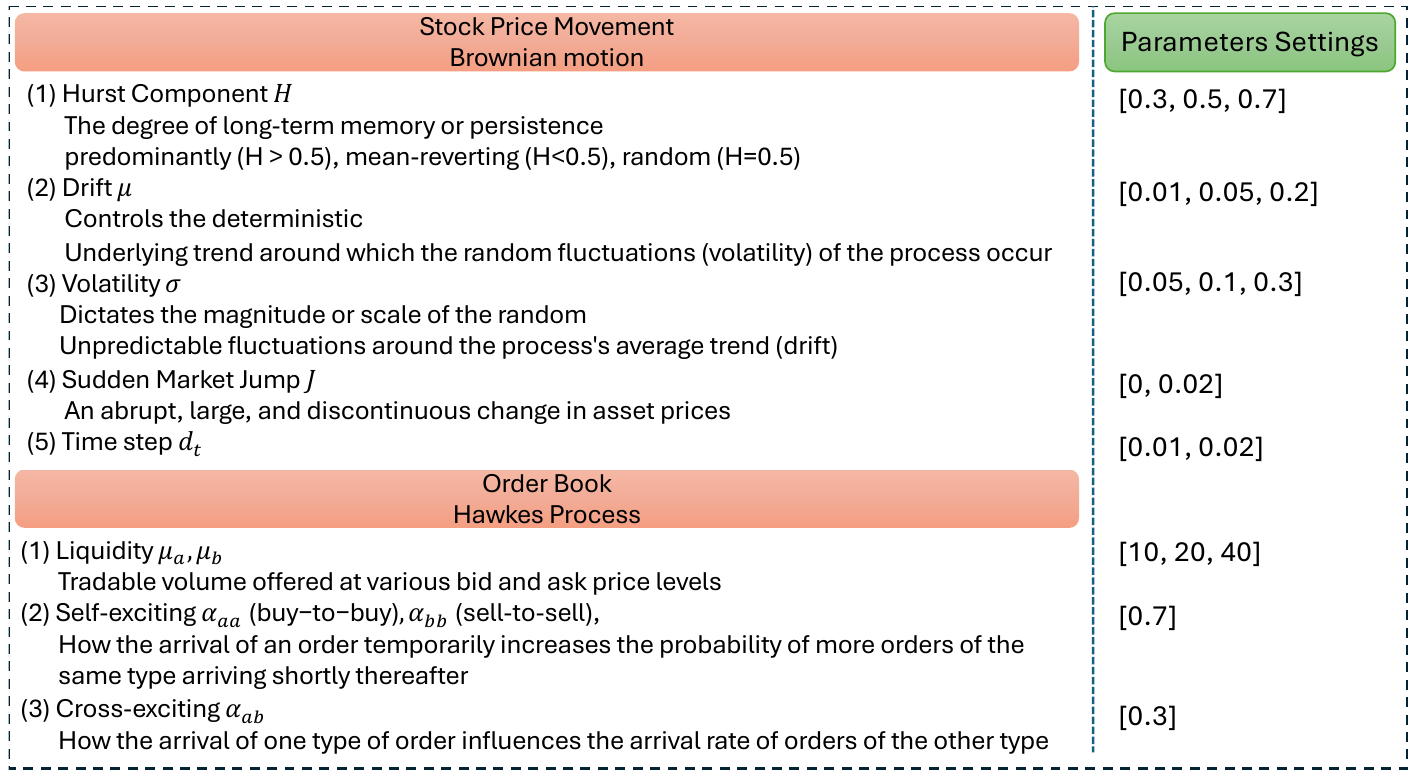}
\end{figure}

\subsection{Generation of State-Action Pairs}   
To generate diverse expert demonstrations for FlowHFT, we utilized a set of expert candidates, including: (1) the Avellaneda-Stoikov (AS) model \cite{avellaneda2008high}, a foundational stochastic control framework for optimal quoting that considers inventory risk and market volatility; (2) the Guéant-Lehalle-Fernandez-Tapia (GLFT) model \cite{gueant2013dealing}, which extends such optimal control approaches, often incorporating aspects such as order flow dynamics; (3) a modified GLFT model incorporating a price drift component, designed to adapt strategies in markets exhibiting directional trends; and (4)a model-free Reinforcement Learning (RL) agent trained with Proximal Policy Optimization (PPO) \cite{schulman2017proximal}, which learns a policy through direct interaction or simulation to maximize a specified reward function. 
The detailed mathematical formulations, assumptions, and parameterization specifics for each of these expert models are provided in Appendix \ref{appendix:Expert Market-Making Models}.

We generate state-action pairs for each market generated. 
For each parameter combination, we simulate $100$ episodes. Each episode consists of $1/dt$ time steps (e.g., $100$ steps if $dt=0.01$), each timestamp yields a state-action pair from the expert strategy, here we collect the corresponding state-action pairs from the dominant stochastic strategy among our candidates(AS, GLFT, GLFT-Drift, PPO). In total, we have gathered 3.24 million state-action pairs. These generated state-action pairs serve as the training dataset for the diffusion policy, providing a broad coverage of market conditions and ensuring robust policy learning. This dataset serves as the training data for our diffusion policy framework.


\section{Experiments and Results Analysis}
\label{sec:Experiments and Results Analysis}
Our investigation is designed to critically evaluate FlowHFT's advantages over existing approaches and its capabilities to earn profit. 
Specifically, we address the following research questions:
\textbf{RQ1:} Can FlowHFT effectively generalize strategies learned from expert demonstrations to new, unseen market conditions? \textbf{RQ2:} Does the fine-tuning mechanism improve the performance of actions initially proposed by the pre-trained model? \textbf{RQ3:} Can the FlowHFT framework achieve greater profitability than traditional strategies?

\subsection{Out-of-sample Market Environment Settings}
To evaluate FlowHFT's performance across a spectrum of out-of-sample market environments, we systematically configured distinct test conditions by setting key market parameters to differ from those used during training (see appendix \ref{appendix: arameters for Test Environemnt} ). Firstly, to simulate varied stock price trends, we controlled the Hurst exponent, which governs the long-term memory and persistence patterns of the price series, and the drift term, which dictates its underlying direction. Secondly, based on these trend characteristics, we further differentiated the market microstructure by creating four specific situations based on combinations of market volatility (Vol) and overall order arrival rate (AR): (1) High Vol/High AR (HH), representing active, potentially news-driven volatile markets; (2) High Vol/Low AR (HL), mimicking risky market; (3) Low Vol/High AR (LH), reflecting stable, liquid markets; and (4) Low Vol/Low AR (LL), simulating quiet market periods. For each of these market situations, we perform traditional strategies as well as FlowHFT to compare their performance.

\subsection{Evaluation Metrics}

\begin{itemize}
    \item \textbf{Profit and Loss (PnL):} Measure the total change in the value over a specific period, reflecting the aggregate percentage gain or loss.
    \item \textbf{Sharpe Ratio (SR):} It helps investors understand how much excess return an investment generated for each unit of risk it undertook. A higher Sharpe Ratio generally indicates a better risk-adjusted performance. 
    \item \textbf{Maximum Drawdown (MDD):} Quantify the largest percentage decline in the value of a strategy from a previous peak to a subsequent trough
\end{itemize}
The detailed calculations of these metrics are provided in Appendix \ref{Calculation of Evaluation Metrics}

\subsection{Results Analysis}
Looking at the results presented from Table~\ref{tab:performance_without_drift_H5} to Table~\ref{tab:performance_with_drift_H8_2}, a consistent pattern emerges across the various market scenarios. Firstly, the GLFT model typically yields superior results compared to the AS model. Critically, our pre-trained flow matching policy model demonstrates performance levels comparable to the GLFT instructor. This finding is significant to address the \textbf{RQ1}. It indicates that the pre-trained model successfully learned and internalized effective strategies from its instructors, even within these challenging stochastic environments. 

In response to \textbf{RQ2}, we examine the results corresponding to the fine-tuned Flow Matching policy. This adapted model typically and significantly outperforms both the traditional benchmark models and the initial pre-trained model across the tested environments. This finding validates the second core objective of our framework: we not only demonstrate the ability to effectively learn from expert instructors but also showcase the capacity to surpass their performance. This success is attributed to the fine-tuning process, which empowers the model to "sense" the prevailing market conditions via the validation set calibration and subsequently adjust its behavior, specializing its strategy for the specific market environment encountered.

Note that our experimental design includes highly volatile market environments, which are characterized by the potential for sudden and significant price movements. We show that FlowHFT exhibits robustness in these challenging conditions, particularly when compared with traditional RL methods. This indicates that generating sequences of actions rather than purely single-step decisions can effectively incorporate a longer-term planning perspective. This approach allows the model to better anticipate downstream consequences, make more considered adjustments over multiple steps, and mitigate the risk of compounding errors.

In response to \textbf{RQ3}, we take analyses of the performance metrics to obtain significant insights into the profit-generating capabilities across different market conditions. The Finetuned Flow Policy consistently demonstrates strong PnL (Profit and Loss) figures, notably achieving the highest PnL of 26.79 in the "Low Volatility \& High Demand" scenario and 25.82 in the "High Volatility \& High Demand" scenario. This indicates its robust ability to capitalize on profit opportunities in both highly active and more stable, liquid markets. Compared to the traditional models, the Finetuned Flow Policy generally matches or surpasses the PnL of AS, GLFT, and GLFT-drift in most scenarios, particularly excelling in high demand situations. The RL-PPO model, while adaptive, consistently shows lower PnL figures across all tested environments, underscoring the superior profit extraction of the FlowHFT framework.

\begin{table}[!htbp]
\centering
\caption{Parameters for Market Conditions:  Hurst Exponent $H=0.5$, Drift Rate $\mu=0$, volatility ($\sigma$: \{0.02, 0.25\}) and arrival rate ($\mu$: \{25, 50\}). The linear finetuning (ax+b)  are the following:
{HH: action+[0.15,0.24], HL: action+[0.1,0.2], LH: action+[-0.1,0.1], LL: action+[-0.3,0.3]}, these parameters can be chosen from an interval, just pick a point as representation
The evaluation metrics are tested out of sample, with each methods are tested 1 million trails to obtain significance and convergence.
}
\renewcommand{\arraystretch}{1.2} 
\resizebox{\textwidth}{!}{%
\begin{tabular}{lrrr|rrr|rrr|rrr}
\toprule
& \multicolumn{3}{c|}{High Volatility \& High Demand} & \multicolumn{3}{c|}{High Volatility \& Low Demand} & \multicolumn{3}{c|}{Low Volatility \& High Demand} & \multicolumn{3}{c}{Low Volatility \& Low Demand} \\
& PnL $\uparrow$ & SR $\uparrow$ & MDD $\downarrow$ & PnL $\uparrow$ & SR $\uparrow$ & MDD $\downarrow$ & PnL $\uparrow$ & SR $\uparrow$ & MDD $\downarrow$ & PnL $\uparrow$ & SR $\uparrow$ & MDD $\downarrow$ \\
\midrule
AS & 24.22& 0.09& 241.65& 13.54& 0.09& 125.78& 25.20 & 1.05 & 7.66 & 13.67 & 0.72 & 6.61 \\
GLFT & 25.10& 0.37& 60.57& 13.56& 0.24& 52.55& 25.87 & 1.17 & 6.95 & 13.91 & 0.78 & 6.14 \\
GLFT-drift & 25.10& 0.37& 60.57& 13.56& 0.24& 52.55& 25.87 & 1.17 & 6.95 & 13.91 & 0.78 & 6.14 \\
RL-PPO & 14.76& 0.10& 133.61& 9.29& 0.08& 103.85& 26.74 & 0.81 & 10.13 & 19.80 & 0.46 & 14.56 \\
\midrule
Pretrained Flow Policy  & 21.99& 0.34& 56.94& 13.28& 0.23& 53.43& 22.29& 1.5& 4.07& 12.10& 0.91& \textbf{4.2}\\
Finetuned Flow Policy& \textbf{25.82}& \textbf{0.38}& \textbf{58.38}& \textbf{14.43}& \textbf{0.26}& \textbf{49.64}& \textbf{26.79}& \textbf{1.62}& \textbf{3.62}& \textbf{22.31}& \textbf{0.96}& 5.17\\
\bottomrule
\end{tabular}%
}

\label{tab:performance_without_drift_H5}
\end{table}

\begin{table}[!htbp]
\centering
\caption{Performance Comparison Under Different Conditions (Hurst Exponent $H=0.2$, Drift Rate $\mu=0.02$).Same measures as the above table. Diffusion Policy select GLFT-drift strategy as the expert to learn from. The linear finetuning (ax+b) for each action is {HH: action+[-0.1,0.1], HL: action+[-0.1,0.1], LH: action+[-0.2,0.2], LL: action+[-0.3,0.3]}, these parameters can be chosen from multiple choices,  just pick one as representation}
\renewcommand{\arraystretch}{1.2} 
\resizebox{\textwidth}{!}{%
\begin{tabular}{lrrr|rrr|rrr|rrr}
\toprule
& \multicolumn{3}{c|}{High Volatility \& High Demand} & \multicolumn{3}{c|}{High Volatility \& Low Demand} & \multicolumn{3}{c|}{Low Volatility \& High Demand} & \multicolumn{3}{c}{Low Volatility \& Low Demand} \\
& PnL $\uparrow$ & SR $\uparrow$ & MDD $\downarrow$ & PnL $\uparrow$ & SR $\uparrow$ & MDD $\downarrow$ & PnL $\uparrow$ & SR $\uparrow$ & MDD $\downarrow$ & PnL $\uparrow$ & SR $\uparrow$ & MDD $\downarrow$ \\
\midrule
GLFT &  29.14&  0.35&  136.07&  15.77&  0.23&  116.71&  29.27&  0.86&  42.06&  16.1&  0.57&  35.81\\
GLFT-drift &  29.96&  0.36&  136.4&  16.53&  0.24&  117.15&  35.23&  0.96&  44.55&  21.2&  0.69&  38.11\\
\midrule
Pretrained Flow Policy &  26.67&  0.35&  \textbf{121.2}&  14.9&  0.21&  \textbf{114.29}&  26.5&  1.11&  \textbf{27.2}&  14.52&  0.67&  \textbf{27.05}\\
Finetuned Flow Policy &  \textbf{31.3}&  \textbf{0.36}&  140.71&  \textbf{18.41}&  \textbf{0.24}&  127.55&  \textbf{38.6}&  \textbf{1.14}&  42.58&  \textbf{28.07}&  \textbf{0.75}&  49.98\\
\bottomrule
\end{tabular}%
}
\label{tab:performance_with_drift_H2_2}
\end{table}

\begin{figure}[htp]
\caption{Scaling analysis showing the relationship between the Number of Instructors (expert models) used in training FlowHFT and its resulting performance, as measured by Sharpe Ratio (blue bars, left y-axis) and Maximum Drawdown (red line, right y-axis).}
\label{scallinglaw}
\centering
\includegraphics[width=0.82\textwidth]{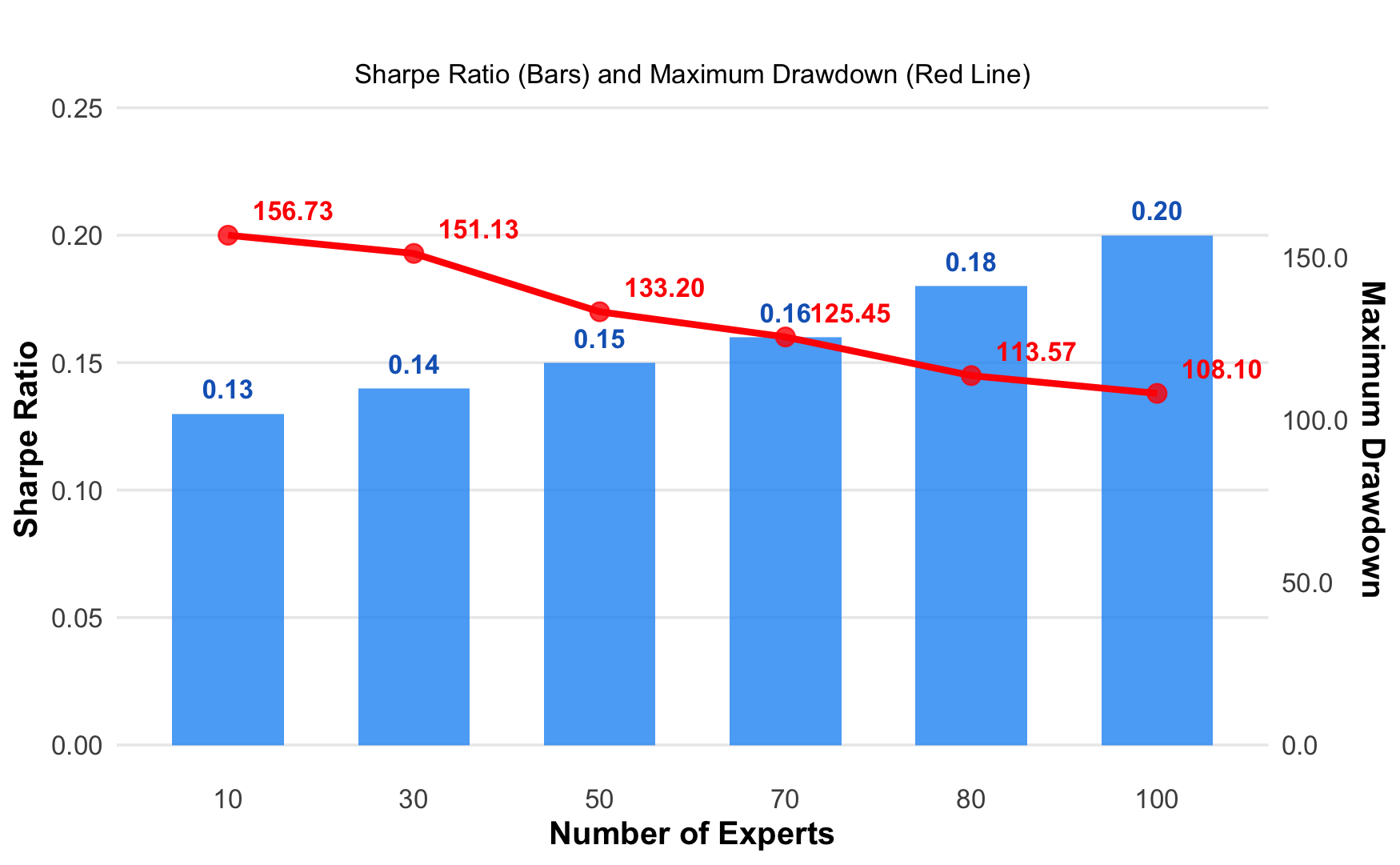}
\end{figure}

The FlowHFT framework, with its flow matching policy, is designed to synthesize knowledge from numerous expert instructors. While our extensive experiments confirm its learning capabilities across many experts, we conduct a analysis to demonstrate its scaling law: that is, how performance improves as the model learns from an increasing number of distinct expert strategies. The results of this scaling analysis are presented in Figure \ref{scallinglaw}.

Figure \ref{scallinglaw} illustrates the scaling properties of our FlowHFT framework, supporting the hypothesis that the model benefits from exposure to an increasing number of expert strategies. Two key observations underscore this scaling law. First, with an increase in the number of expert instructors used during training, the framework's performance consistently improves over configurations with fewer experts. This progressive improvement indicates that FlowHFT effectively learns and integrates new, valuable strategic information from additional experts. Second, this enhanced learning translates into more robust risk-adjusted returns; not only does the Sharpe Ratio increase, but the Maximum Drawdown (red line) simultaneously shows a general decline. This continuous improvement in both return and risk metrics as more expert knowledge is incorporated suggests that FlowHFT is adept at leveraging a broader set of specialized strategies to find more consistently superior actions across diverse market conditions.


\section{Conclusion}

In conclusion, we propose a novel imitation learning framework FlowHFT, the first flow matching policy-based approach in a financial stochastic control task. Our findings confirm that it has ability to learn effective trading strategies in various stochastic environments in one model, thus enabling prompt adaptation to prevailing market conditions. Furthermore, the incorporated grid search mechanism successfully refines actions, especially in situations where these experts were suboptimal. 
In terms of profit generation, FlowHFT's advantage over traditional models is evident through its achievement of greater cumulative returns, a superior Sharpe ratio indicating better risk-adjusted performance, and a reduced maximum drawdown signifying lower downside risk.
Therefore, FlowHFT represents a promising advancement in the development of adaptive and high-performance HFT strategies.








\bibliographystyle{plain}  
\bibliography{references}

\appendix

\section{Expert Market-Making Models}
\label{appendix:Expert Market-Making Models}

\subsection{Avellaneda-Stoikov (AS) Model}
The Avellaneda-Stoikov (AS) model~\cite{avellaneda2008high} optimizes bid and ask spreads by maximizing the expected utility of terminal wealth. Taking the coefficients of drift, jump intensity,  self-excitation and mutual-excitation to be zero in above settings , The value function $u(s, x, q, t)$ is governed by the Hamilton-Jacobi-Bellman (HJB) equation (details in Appendix):

\begin{align}
\frac{\partial u}{\partial t} + \frac{1}{2} \sigma^2 \frac{\partial^2 u}{\partial s^2} 
&+ \max_{\delta_b} \lambda_b\left(\delta_b\right)\left[u\left(s, x-s+\delta_b, q+1, t\right)-u(s, x, q, t)\right] \notag \\
&+ \max_{\delta_a} \lambda_a\left(\delta_a\right)\left[u\left(s, x+s+\delta_a, q-1, t\right)-u(s, x, q, t)\right] = 0, \label{eq:HJB_AS}
\end{align}

with terminal condition $u(s, x, q, T) = -e^{-\gamma\left(x + q S_T\right)}$.

The optimal bid and ask spreads are:

\begin{align}
\delta_{\text{bid}}^{\text{AS}} &= \frac{\gamma \sigma^2 \tau}{2} + \frac{1}{\gamma} \ln\left(1 + \frac{\gamma}{k}\right) - q \gamma \sigma^2 \tau, \label{eq:delta_bid_AS} \\
\delta_{\text{ask}}^{\text{AS}} &= \frac{\gamma \sigma^2 \tau}{2} + \frac{1}{\gamma} \ln\left(1 + \frac{\gamma}{k}\right) + q \gamma \sigma^2 \tau, \label{eq:delta_ask_AS}
\end{align}



In the model, $\gamma$ is defined as the risk aversion parameter, which quantifies the trader's preference for avoiding risk relative to seeking returns. The variable $\sigma$ represents the asset's volatility, indicating the degree of price fluctuation over time. The term $\tau = T - t$ denotes the remaining time horizon, calculating the time left until the end of the trading period. The parameter $k$ measures the sensitivity of order arrivals to price spreads, demonstrating how spread changes influence trading activity. Lastly, $q$ is the current inventory level, reflecting the amount of the asset held by the trader.

\subsection{Guéant-Lehalle-Fernandez-Tapia (GLFT) Model}
The Guéant-Lehalle-Fernandez-Tapia (GLFT) model~\cite{gueant2013dealing} offers an approximate solution for optimal spreads by incorporating additional terms into the HJB equation. The assumptions are keeping the same as AS model expect. There is no limitation to the terminal time T. The optimal spreads are:

\begin{align}
    c_1 &= \frac{1}{\gamma} \ln\left(1 + \frac{\gamma}{k}\right), \\
    c_2 &= \sqrt{\frac{\gamma}{2 A k} \left(1 + \frac{\gamma}{k}\right)^{\left(\frac{k}{\gamma} + 1\right)}}, \\
    \delta_{\text{bid}}^{\text{GLFT}} &= c_1 + \frac{\sigma c_2}{2} + \sigma c_2 q, \\
    \delta_{\text{ask}}^{\text{GLFT}} &= c_1 + \frac{\sigma c_2}{2} - \sigma c_2 q, \\
\end{align}




In the model, $A$ denotes the base arrival rate of orders. The coefficients $c_1$ and $c_2$ are used to adjust the bid and ask spreads based on factors such as risk aversion, volatility, and time horizon. Lastly, $q$ represents the current inventory level, indicating the amount of the asset that is held.

\subsection{Modified Guéant-Lehalle-Fernandez-Tapia (GLFT) Model with Drift}
Keeping everything unchanged except for the drift part, the new modified GLFT solution which hasn't been shown in the original paper\cite{gueant2013dealing}.
\[|q| = Q_{\max}\]

\[dS_t = \mu dt + \sigma dW_t\]

\[(\delta^b_{\infty})^* = \frac{1}{\gamma}\ln(1 + \frac{\gamma}{k}) + \left[-\frac{\mu}{\gamma\sigma^2} + \frac{2q+1}{2}\right]\sqrt{\frac{\sigma^2}{2kA}(1 + \frac{\gamma}{k})^{1+\frac{k}{\gamma}}}\]

\[(\delta^a_{\infty})^* = \frac{1}{\gamma}\ln(1 + \frac{\gamma}{k}) + \left[\frac{\mu}{\gamma\sigma^2} - \frac{2q-1}{2}\right]\sqrt{\frac{\sigma^2}{2kA}(1 + \frac{\gamma}{k})^{1+\frac{k}{\gamma}}}\]

\subsection{Proximal Policy Optimization (PPO) for High-Frequency Market Making}
\label{sec:PPO}

To apply Proximal Policy Optimization (PPO) in a high-frequency trading (HFT) environment, we formulate the trading process as a Markov Decision Process (MDP). The input to the policy model consists of high-dimensional market state features, such as limit order book snapshots, recent trade volumes, price trends, and microstructure indicators sampled at millisecond intervals. The action space can include discrete actions like placing a buy/sell/cancel order at a specific price level. The reward function is designed to capture profit-and-loss (PnL) adjusted for market impact and inventory risk, including components like realized spread, execution cost, and inventory penalties. The objective is to train a policy that maximizes expected cumulative rewards.

In the context of market making, the RL framework can be formalized as a Markov Decision Process (MDP) defined by the tuple $(\mathcal{S}, \mathcal{A}, \mathcal{P}, \mathcal{R}, \gamma)$, 
where, $S$ represents the state space, and $A$ denotes the action space. The state transition probabilities are symbolized by $P$, which dictates the likelihood of transitioning from one state to another given a particular action. $R$ is the reward function, which assigns a reward based on the state and action taken. Finally, $\gamma \in [0, 1]$ is the discount factor, used to determine the present value of future rewards, allowing for the evaluation of long-term versus immediate benefits.

For the space state, at each discrete time step $t$, the state $s_t$ encapsulates essential information required for decision-making:
\[
s_t = (t, x_t, q_t, S_t, B_{t-1}-A_{t-1}),
\]

In the model, $t$ is the timestamp indicating the specific point in time under consideration. The variable $x_t$ denotes the current cash position, while $q_t$ represents the current inventory, which refers to the number of shares held. The term $S_t$ is the current underlying stock price, providing a snapshot of the stock's market value at time $t$. $A_{t-1} - B_{t-1}$ is the difference between the ask and bid quotes from the previous time step, reflecting the spread.

For the action space, the agent's action $a_t$ consists of setting new ask and bid quotes relative to the mid-price $M_t$. Specifically:
\[
a_t = (\delta^{\text{ask}}_t, \delta^{\text{bid}}_t),
\]

In the model, $\delta_{\text{ask}t}$ represents the offset of the ask quote from the mid-price, which sets the ask price to $M_t + \delta{\text{ask}t}$. Conversely, $\delta{\text{bid}t}$ is the offset of the bid quote from the mid-price, setting the bid price to $M_t - \delta{\text{bid}_t}$.

For the reward function, it is designed to balance profitability from the bid-ask spread against the risk associated with inventory holding. The formulation is:
\[
r_t = \text{PnL}_t - \lambda q_t^2,
\]
$PnL_t$ represents the profit and loss at time $t$, derived from executed trades. The parameter $\lambda$ is a risk aversion parameter that penalizes large inventory positions.

After defining the MDP, policy gradient methods is used to optimize the policy $\pi_\theta(a|s)$ parameterized by $\theta$ to maximize the expected cumulative reward:
\[
J(\theta) = \mathbb{E}_{\tau \sim \pi_\theta} \left[ \sum_{t=0}^{T} \gamma^t r_t \right],
\]
where $\tau = (s_0, a_0, s_1, a_1, \ldots, s_T)$ denotes a trajectory sampled from the policy.
The output of the policy network is a parameterized probability distribution over actions, from which the actual trading action is sampled.
The derivation related to policy gradient is in Appendix.

\section{Calculation of Evaluation Metrics}
\label{Calculation of Evaluation Metrics}
\subsection*{1. Cumulative Return (CR)}
The Cumulative Return measures the total percentage change in the value of an investment over a specific period.


If $r_i$ is the return for period $i$, and there are $n$ periods:
\[
\text{CR} = \left( \prod_{i=1}^{n} (1 + r_i) - 1 \right) \times 100\%
\]
where:
\begin{itemize}
    \item $r_i$: Return for period $i$.
    \item $n$: Total number of periods.
\end{itemize}

\subsection*{2. Sharpe Ratio (SR)}
The Sharpe Ratio measures the risk-adjusted return of an investment or strategy. It quantifies the average return earned in excess of the risk-free rate per unit of volatility or total risk.
\[
\text{SR} = \frac{\mathbb{E}[R_p - R_f]}{\sigma_{p}}
\]

where:
\begin{itemize}
    \item $\mathbb{E}[R_p - R_f]$: The expected value of the excess return of the portfolio over the risk-free rate.
    \item $\bar{R}_p$: The average (sample mean) return of the portfolio or investment over a period.
    \item $\bar{R}_f$: The average (sample mean) risk-free rate of return over the same period (e.g., yield on short-term government bonds).
    \item $\sigma_p$: The standard deviation of the portfolio's excess returns ($R_p - R_f$). (Sometimes, the standard deviation of the portfolio's returns, $\sigma(R_p)$, is used as an approximation, especially if $R_f$ is constant or has very low volatility).
\end{itemize}

\subsection*{3. Maximum Drawdown (MDD)}
The Maximum Drawdown is the largest percentage drop in the value of an investment or portfolio from a previous peak to a subsequent trough over a specific period. Let $V(t)$ be the value of the portfolio at time $t$ over the period $[0, T]$.

First, define the running peak value up to time $t$ as $M(t)$:
\[
M(t) = \max_{0 \le \tau \le t} V(\tau)
\]
Then, the drawdown $D(t)$ at any time $t$ is:
\[
D(t) = \frac{M(t) - V(t)}{M(t)}
\]
The Maximum Drawdown (MDD) is the maximum value of $D(t)$ over the entire period:
\[
\text{MDD} = \max_{0 \le t \le T} D(t)
\]
This MDD is usually expressed as a positive percentage (e.g., an MDD of 0.25 means a 25\% maximum loss from a peak).

where:
\begin{itemize}
    \item $V(t)$: Value of the investment at time $t$.
    \item $T$: The end of the observation period.
    \item $\tau$: A time index variable ranging from $0$ to $t$.
\end{itemize}

\section{Parameters for Test Environemnt}
\label{appendix: arameters for Test Environemnt}
\begin{figure}[htp]
\caption{Parameters settings of testing market environement}
\label{}
\centering
\includegraphics[width=0.82\textwidth]{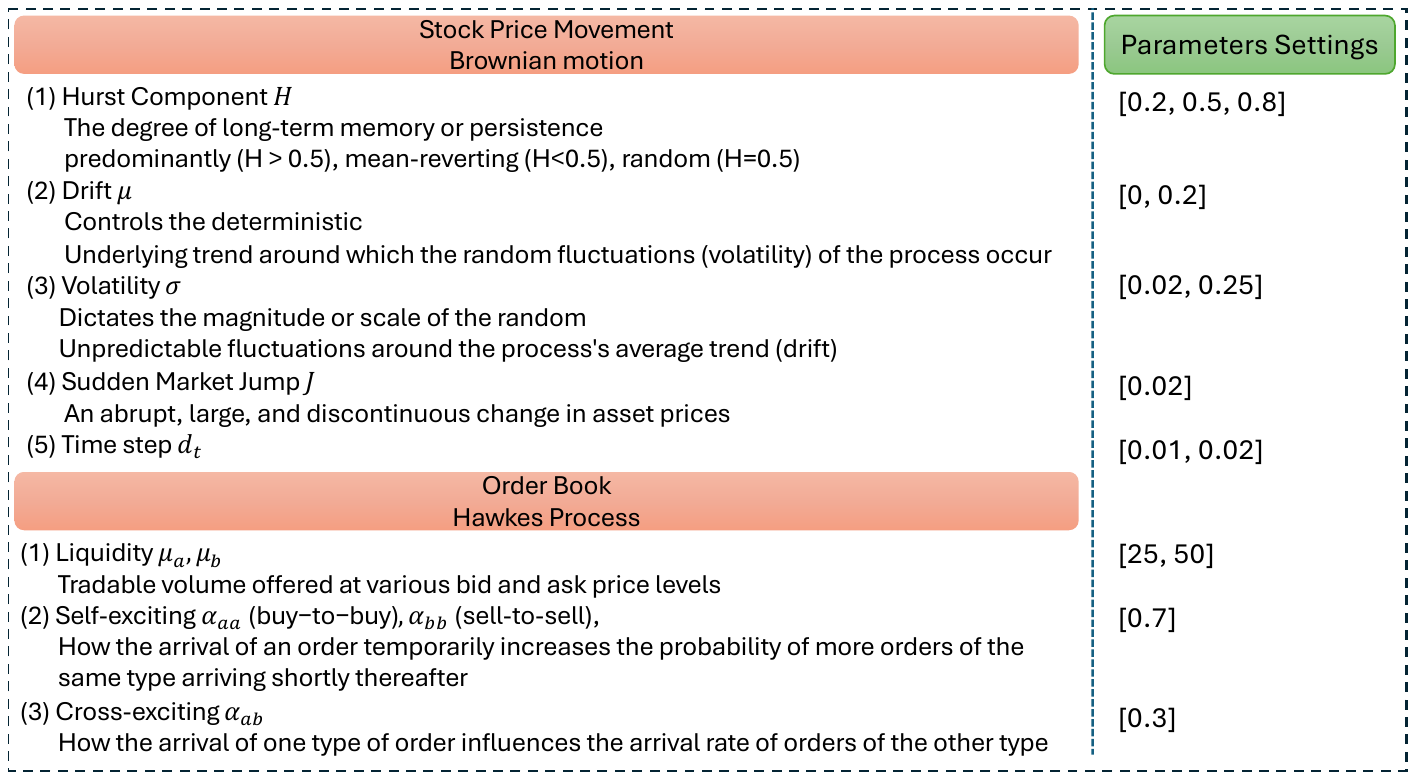}
\end{figure}

\section{Results for Different Market Conditions}

\begin{table}[!htbp]
\centering
\caption{Performance Comparison Under Different Conditions (Hurst Exponent $H=0.5$, Drift Rate $\mu=0.02$).Same measures as the above table. Flow Policy select GLFT-drift strategy as the expert to learn from. The linear finetuning (ax+b) for each action is {HH: action+[0.1,0.25], HL: action+[0.1,0.2], LH: action+[-0.3,0.3], LL: action+[-0.4,0.7]}, these parameters can be chosen from multiple choices,  just pick one as representation}
\renewcommand{\arraystretch}{1.2} 
\resizebox{\textwidth}{!}{%
\begin{tabular}{lrrr|rrr|rrr|rrr}
\toprule
& \multicolumn{3}{c|}{High Volatility \& High Demand} & \multicolumn{3}{c|}{High Volatility \& Low Demand} & \multicolumn{3}{c|}{Low Volatility \& High Demand} & \multicolumn{3}{c}{Low Volatility \& Low Demand} \\
& PnL $\uparrow$ & SR $\uparrow$ & MDD $\downarrow$ & PnL $\uparrow$ & SR $\uparrow$ & MDD $\downarrow$ & PnL $\uparrow$ & SR $\uparrow$ & MDD $\downarrow$ & PnL $\uparrow$ & SR $\uparrow$ & MDD $\downarrow$ \\
\midrule
AS &  24.16&  0.08&  245.49&  13.51&  0.09&  127.70&  25.18&  0.92&  8.02&  13.61&  0.63&  6.98\\
GLFT &  25.11&  0.36&  61.39&  13.55&  0.23&  53.26&  25.21&  0.99&  7.45&  13.62&  0.66&  6.59\\
GLFT-drift &  25.87&  0.37&  61.26&  14.15&  0.24&  53.21&  40.82&  1.10&  7.69&  26.44&  0.87&  6.97\\
RL-PPO &  30.74&  0.16&  186.1&  \textbf{20.47}&  0.14&  138.88&  24.79&  0.72&  10.58&  \textbf{32.49}&  0.65&  15.28\\
\midrule
Pretrained Flow Policy &  22.72&  0.35&  \textbf{57.09}&  13.28&  0.23&  53.43&  21.90&  \textbf{1.64}&  \textbf{3.97}&  11.96&  \textbf{0.98}&  \textbf{4.05}\\
Finetuned Flow Policy &  \textbf{27.45}&  \textbf{0.39}&  61.31&  15.15&  \textbf{0.26}&  \textbf{50.95}&  \textbf{42.81}&  1.17&  8.44&  27.12&  0.93&  7.42\\
\bottomrule
\end{tabular}%
}
\label{tab:performance_with_drift_H5}
\end{table}

\newpage
Tables 4 to 6 expand the volatility and arrival rate options to investigate a broader spectrum of unseen scenarios. (Volatility, $\sigma$: \{0.05, 0.2\}) and arrival rates, $\mu$: \{30, 60\})
\begin{table}[!htbp]
\centering
\caption{Performance Comparison Under Different Conditions (Hurst Exponent $H=0.2$, Drift Rate $\mu=0.0$). Same measures as the above table. Diffusion Policy select GLFT-drift strategy as the expert to learn from. The linear finetuning (ax+b) for each action is {HH: action+[0.02,0.2], HL: action+[0.05,0.15], LH: action+[-0.15,0.15], LL: action+[-0.1,0.1]}, these parameters can be chosen from multiple choices,  just pick one as representation }
\renewcommand{\arraystretch}{1.2} 
\resizebox{\textwidth}{!}{%
\begin{tabular}{lrrr|rrr|rrr|rrr}
\toprule
& \multicolumn{3}{c|}{High Volatility \& High Demand} & \multicolumn{3}{c|}{High Volatility \& Low Demand} & \multicolumn{3}{c|}{Low Volatility \& High Demand} & \multicolumn{3}{c}{Low Volatility \& Low Demand} \\
& PnL $\uparrow$ & SR $\uparrow$ & MDD $\downarrow$ & PnL $\uparrow$ & SR $\uparrow$ & MDD $\downarrow$ & PnL $\uparrow$ & SR $\uparrow$ & MDD $\downarrow$ & PnL $\uparrow$ & SR $\uparrow$ & MDD $\downarrow$ \\
\midrule
AS &  28.82&  0.14&  346.8&  16.23&  0.13&  216.69&  29.28&  0.78&  48.86&  16.16&  0.54&  40.22\\
GLFT &  29.16&  0.36&  134.33&  15.78&  0.23&  115.22&  29.29&  0.92&  41.34&  16.1&  0.61&  35.2\\
GLFT-drift &  29.16&  0.36&  134.33&  15.78&  0.23&  115.22&  29.29&  0.92&  41.34&  16.1&  0.61&  35.2\\
\midrule
Pretrained Flow Policy &  26.61&  0.35&  \textbf{119.73}&  14.76&  0.22&  \textbf{112.78}&  26.56&  \textbf{1.16}&  \textbf{26.76}&  14.61&  0.71&  \textbf{26.58}\\
Finetuned Flow Policy &  \textbf{29.72}&  \textbf{0.36}&  136.52&  \textbf{15.80}&  \textbf{0.23}&  114.62&  \textbf{31.43}&  1.06&  36.25&  \textbf{16.82}&  \textbf{0.75}&  29.33\\
\bottomrule
\end{tabular}%
}
\label{tab:performance_without_drift_H2_0}
\end{table}

\newpage
\begin{table}[!htbp]
\centering
\caption{Performance Comparison Under Different Conditions (Hurst Exponent $H=0.8$, Drift Rate $\mu=0.0$).Same measures as the above table. Diffusion Policy select GLFT-drift strategy as the expert to learn from. The linear finetuning (ax+b) for each action is {HH: action+[-0.02, 0.1], HL: action+[0.1, 0.1], LH: action+[-0.1, 0.1], LL: action+[-0.1, 0.1]}, these parameters can be chosen from multiple choices,  just pick one as representation}
\renewcommand{\arraystretch}{1.2} 
\resizebox{\textwidth}{!}{%
\begin{tabular}{lrrr|rrr|rrr|rrr}
\toprule
& \multicolumn{3}{c|}{High Volatility \& High Demand} & \multicolumn{3}{c|}{High Volatility \& Low Demand} & \multicolumn{3}{c|}{Low Volatility \& High Demand} & \multicolumn{3}{c}{Low Volatility \& Low Demand} \\
& PnL $\uparrow$ & SR $\uparrow$ & MDD $\downarrow$ & PnL $\uparrow$ & SR $\uparrow$ & MDD $\downarrow$ & PnL $\uparrow$ & SR $\uparrow$ & MDD $\downarrow$ & PnL $\uparrow$ & SR $\uparrow$ & MDD $\downarrow$ \\
\midrule
GLFT &  29.00&  0.57&  21.05&  15.77&  0.36&  19.88&  29.21&  1.10&  6.63&  16.07&  0.73&  6.38\\
GLFT-drift &  29.00&  0.57&  21.05&  15.77&  0.36&  19.88&  29.21&  1.10&  6.63&  16.07&  0.73&  6.38\\
\midrule
Pretrained Flow Policy &  26.33&  0.67&  \textbf{17.37}&  14.53&  0.36&  \textbf{19.26}&  26.61&  \textbf{1.66}&  3.80&  14.45&  0.93&  4.57\\
Finetuned Flow Policy &  \textbf{30.31}&  \textbf{0.67}&  17.77&  \textbf{17.61}&  \textbf{0.39}&  19.66&  \textbf{29.59}&  1.50&  \textbf{3.42}&  \textbf{16.65}&  \textbf{0.96}&  \textbf{4.22}\\
\bottomrule
\end{tabular}%
}
\label{tab:performance_without_drift_H8_0}
\end{table}

\begin{table}[!htbp]
\centering
\caption{Performance Comparison Under Different Conditions (Hurst Exponent $H=0.8$, Drift Rate $\mu=0.02$).Same measures as the above table. Diffusion Policy select GLFT-drift strategy as the expert to learn from. The linear finetuning (ax+b) for each action is {HH: action+[-0.1, 0.1], HL: action+[-0.1, 0.1], LH: action+[-0.1, 0.1], LL: action+[-0.1, 0.1]}, these parameters can be chosen from multiple choices,  just pick one as representation}
\renewcommand{\arraystretch}{1.2} 
\resizebox{\textwidth}{!}{%
\begin{tabular}{lrrr|rrr|rrr|rrr}
\toprule
& \multicolumn{3}{c|}{High Volatility \& High Demand} & \multicolumn{3}{c|}{High Volatility \& Low Demand} & \multicolumn{3}{c|}{Low Volatility \& High Demand} & \multicolumn{3}{c}{Low Volatility \& Low Demand} \\
& PnL $\uparrow$ & SR $\uparrow$ & MDD $\downarrow$ & PnL $\uparrow$ & SR $\uparrow$ & MDD $\downarrow$ & PnL $\uparrow$ & SR $\uparrow$ & MDD $\downarrow$ & PnL $\uparrow$ & SR $\uparrow$ & MDD $\downarrow$ \\
\midrule
GLFT &  28.98&  0.56&  21.46&  15.76&  0.35&  20.23&  29.19&  1.00&  6.95&  16.07&  0.67&  6.68\\
GLFT-drift &  30.05&  0.57&  21.08&  16.74&  0.37&  19.94&  \textbf{35.22}&  1.11&  5.71&  \textbf{21.24}&  0.80&  5.60\\
\midrule
Pretrained Flow Policy &  26.64&  \textbf{0.66}&  \textbf{17.57}&  14.52&  0.34&  \textbf{19.47}&  26.74&  1.61&  3.79&  14.50&  0.87&  4.68\\
Finetuned Flow Policy &  \textbf{33.17}&  0.62&  19.64&  \textbf{19.20}&  \textbf{0.38}&  20.61&  32.57&  \textbf{1.61}&  \textbf{3.01}&  19.02&  \textbf{1.01}&  \textbf{3.80}\\
\bottomrule
\end{tabular}%
}
\label{tab:performance_with_drift_H8_2}
\end{table}

\section{Limitations}
\label{Limitations}
While FlowHFT demonstrates promising capabilities, certain limitations in the current study should be acknowledged. One limitation of the current study is its focused scope, with the FlowHFT framework being developed and evaluated exclusively for a single financial asset trading on a single exchange. While this controlled setting allows for a detailed analysis of the framework's core mechanics and learning capabilities, we can further address the broader complexities and rich dynamics present in multi-asset or multi-exchange HFT operations. For instance, this study can explore how FlowHFT would manage inter-asset correlations, portfolio-level risk diversification, or the strategic allocation of capital across different instruments. Similarly, the challenges and opportunities arising from operating across multiple exchanges—such as navigating varying fee structures, latency differences, diverse order book microstructures, and potential inter-exchange arbitrage—remain outside the current investigation. Therefore, a significant and crucial avenue for future research is the extension and evaluation of the FlowHFT framework to multi-asset and multi-exchange environments. This would involve adapting the state and action representations, potentially incorporating mechanisms to learn cross-asset dependencies, and developing policies capable of sophisticated order routing and execution across different trading venues. Successfully addressing these aspects would substantially broaden FlowHFT's applicability and provide a more comprehensive validation of its potential as a robust solution for real-world high-frequency trading.

\section{Related Work}
\subsection{Traditional Models for High Frequency Trading}
High-Frequency Trading (HFT) is a subset of algorithmic trading characterized by exceptionally high speeds of order execution and substantial turnover rates. Market making involves the continuous placement of limit orders—both bids (buy orders) and asks (sell orders)—aimed at capturing the bid-ask spread, which is the difference between the highest price a buyer is willing to pay and the lowest price a seller is willing to accept \cite{law2019market, korajczyk2019high}. Traditional models typically formulate the market maker's decision-making process as a stochastic optimal control problem, aiming to devise quoting strategies that maximize predefined objectives (e.g., expected utility of profits) while managing associated risks (e.g., inventory risk) \cite{fodra2012high}.

The Avellaneda-Stoikov (AS) model \cite{avellaneda2008high} is a prominent example within HFT literature. This optimization approach is formulated as maximizing the expected utility of the market maker's terminal wealth at a predetermined finite terminal time. Several key assumptions underpin the AS model: the mid-price of the asset follows an arithmetic Brownian motion without drift \cite{morters2010brownian}; the arrival of market buy and sell orders is modeled as independent Poisson processes \cite{last2017lectures}; and the market maker continuously adjusts bid and ask quotes based on changes in the mid-price and inventory levels, without incurring explicit updating or cancellation costs. Moreover, the AS model often assumes liquidation of inventory (achieving a zero inventory position) by a specified terminal time T, typically the end of the trading day—an assumption potentially unsuitable for assets traded continuously, such as foreign exchange or cryptocurrencies.

The Guéant-Lehalle-Fernandez-Tapia (GLFT) model \cite{gueant2013dealing} represents another widely adopted traditional approach. A central motivation behind its development was to address certain practical and theoretical limitations of the AS model. Notably, the GLFT model does not depend explicitly on a predetermined finite terminal time, making it more suitable for continuously traded assets or markets operating over extended periods. However, a fundamental challenge shared with the AS model is the reliance on historical market data for calibrating key parameters, such as trading intensity coefficients and volatility \cite{gueant2016financial}. Such calibration processes are practically challenging. Empirical evidence also suggests that the exponential form of the trading intensity function may inadequately capture complex relationships between quote depth and execution probability across varying market conditions or price levels. Even if parameters initially derived from the GLFT model are optimal, they tend to become suboptimal quickly as market conditions evolve, necessitating frequent recalibrations—an approach both computationally intensive and susceptible to estimation errors \cite{platt2016problem}.

\subsection{Reinforcement Learning Techniques in High Frequency Trading}
Reinforcement learning (RL) enables an agent to learn rapid, sequential trading decisions by continuously interacting with the market environment \cite{kaelbling1996reinforcement}. The appeal of RL in HFT stems from its inherent capability to learn adaptive strategies within dynamic, complex financial environments \cite{briola2021deep}.

The application of RL to HFT typically involves formulating the trading problem as a Markov Decision Process (MDP) \cite{puterman1990markov}. Under this formulation, the HFT environment (e.g., financial market or order book) evolves through a series of states. The RL agent, representing the trading algorithm, observes the current market state and selects an action (such as placing buy orders, canceling existing orders, or holding positions). Subsequently, the environment transitions to a new state, providing the agent with a numerical reward or penalty reflecting the immediate outcome of its action (e.g., profit or loss). The ultimate objective of the RL agent is to derive an optimal policy that specifies the best action to take in any given state, maximizing the expected cumulative discounted reward over time \cite{ding2020introduction}.

A variety of RL algorithms have been explored for HFT. Q-learning and its deep learning extension, Deep Q-Networks (DQN), along with variants such as Double DQN (DDQN), are widely employed for optimal execution and adaptive trading strategies, as they help reduce overestimation bias \cite{qin2024earnhft, briola2021deep, zhang2022reinforcement}. Actor-Critic methods—which combine policy-based and value-based learning—are also extensively used. Examples include Advantage Actor-Critic (A2C/A3C), enabling parallel learning \cite{zhang2019deep}; Deep Deterministic Policy Gradient (DDPG), well-suited to continuous action spaces like setting order prices or sizes \cite{jeon2024frequant}; and Soft Actor-Critic (SAC), which enhances exploration through entropy regularization \cite{narang2013inside}. Proximal Policy Optimization (PPO) is preferred for its stability and sample efficiency, particularly valuable in dynamic market-making scenarios \cite{zong2024macrohft}. To capture temporal dependencies inherent in financial time series, Recurrent Reinforcement Learning (RRL) is used, while Nash Q-learning addresses multi-agent interactions common in competitive market-making environments \cite{zheng2024reinforcement}.

Despite the advantages, RL strategies in finance are prone to instability. Financial markets are inherently non-stationary; their statistical properties—including mean, variance, and correlations among assets or indicators—change over time. Patterns and relationships valid in one period may fail to generalize subsequently. Consequently, an RL agent trained on historical data representing a particular market regime (e.g., low-volatility trending market) may perform poorly or fail when conditions shift to a new regime (e.g., high-volatility range-bound market).

\subsection{Flow Matching Model in Control}
Flow Matching (FM) has recently emerged as a compelling paradigm in generative modeling, providing an innovative approach to learning complex data distributions \cite{lipman2022flow, dao2023flow}. FM learns transformations that map samples from a simple, well-defined prior distribution (typically an isotropic Gaussian, often termed "noise") to emulate complex target distributions.

Flow matching has been widely adopted in robotics, particularly for generating smooth, continuous control actions for complex tasks \cite{lipman2024flow,zhang2025flowpolicy}. Its primary benefit is the ability to learn velocity fields that directly correspond to how robot actions should evolve over time. For example, tasks involving manipulation or humanoid locomotion benefit from FM's capability to produce sequences of joint movements that are inherently smooth and temporally consistent. This results in predictable and reliable robotic behaviors, facilitating effective interactions with physical environments. Additionally, conditioning flows on inputs like sensor data or task objectives allows robots to develop versatile policies adaptable to changing circumstances while preserving smooth trajectories.

The motivation to adapt flow matching from robotic control to financial markets, specifically HFT, stems from shared challenges: both domains require sophisticated and continuous sequences of actions responsive to high-dimensional, rapidly changing environments. In robotics, FM excels at generating smooth, multi-modal trajectories by learning velocity fields to guide robot actions over time. In HFT, analogous "actions" are buy/sell orders, and the "environment" is the continuously evolving market data.

However, applying flow matching to real-world HFT introduces a critical challenge: transaction speed. Solving the underlying ordinary differential equations (ODEs) usually requires numerous iterative steps, resulting in significant computational costs. To address this, shortcut policies \cite{frans2024one} provide an effective solution by enabling the inference process to execute within a small, fixed number of steps—greatly reducing latency.

\section{Ablation Study}

To validate that our FlowHFT framework indeed learns from the provided expert strategies, we conducted an ablation study. We train FlowHFT model to imitate different sources: (1) purely random actions, (2) only the Avellaneda-Stoikov (AS) model, and (3) only the Guéant-Lehalle-Fernandez-Tapia (GLFT) model. 
The performance of each resulting imitation learning model was subsequently benchmarked against its corresponding source expert—namely, the original traditional AS and GLFT models.

Our results provide clear evidence that FlowHFT successfully learns strategies from expert demonstrations. The model trained on random actions demonstrated the poorest performance, yielding significantly lower PnL and Sharpe Ratios, thus confirming that it failed to learn any meaningful trading strategy. In contrast, the FlowHFT model trained on the AS expert exhibited significantly improved performance compared to the random baseline, achieving metrics comparable to those of the original AS model under identical market conditions. This finding strongly suggests successful imitation of the AS trading strategy. Furthermore, the FlowHFT model trained using the GLFT demonstrations outperformed the AS-imitating model, closely approaching the performance of the original GLFT expert. Considering that the GLFT model is explicitly designed to address certain limitations inherent in the AS model and typically yields superior performance, this outcome provides compelling evidence that FlowHFT effectively captures and replicates the sophisticated strategies of advanced expert models.

\begin{table}[!htbp]
  \centering
  \caption{Parameters for Market Conditions: Hurst Exponent $H=0.5$, Drift Rate $\mu=0$, volatility ($\sigma$: \{0.02, 0.25\}) and arrival rate ($\mu$: \{25, 50\}). 
  }
  \renewcommand{\arraystretch}{1.2}
  \resizebox{\textwidth}{!}{%
  \begin{tabular}{lrrr|rrr|rrr|rrr}
  \toprule
  & \multicolumn{3}{c|}{High Vol. \& High Demand (HH)} & \multicolumn{3}{c|}{High Vol. \& Low Demand (HL)} & \multicolumn{3}{c|}{Low Vol. \& High Demand (LH)} & \multicolumn{3}{c}{Low Vol. \& Low Demand (LL)} \\
  & PnL $\uparrow$ & SR $\uparrow$ & MDD $\downarrow$ & PnL $\uparrow$ & SR $\uparrow$ & MDD $\downarrow$ & PnL $\uparrow$ & SR $\uparrow$ & MDD $\downarrow$ & PnL $\uparrow$ & SR $\uparrow$ & MDD $\downarrow$ \\
  \midrule  
  Random Action & 2.13& 0.07& 28.45& 1.09 & 0.05& 19.25& 2.13 & 0.31 & 2.7  & 1.1 & 0.22 & 1.86 \\
  AS Expert & 24.22& 0.09& 241.65& 13.54& 0.09& 125.78& 25.20 & 1.05 & 7.66 & 13.67 & 0.72 & 6.61 \\
  GLFT-drift Expert& 25.10& 0.37& 60.57& 13.56& 0.24& 52.55& 25.87 & 1.17 & 6.95 & 13.91 & 0.78 & 6.14 \\
  \midrule
  Learning Random Action & -1.06& -0.02& 52.25& -1.01& -0.02& 37.77& -0.76 & -0.06 & 6.00 & -0.50 & -0.06 & 4.24 \\
  Learning AS Only & 22.09& 0.12& 157.54& 11.95& 0.10& 107.11& 22.88 & 0.62 & 13.25 & 12.55 & 0.51 & 9.06 \\
  Learning GLFT-Drift Only & 22.62& 0.36& 58.57& 12.45& 0.20& 57.15& 23.02 & 1.63 & 4.19 & 12.39 & 0.98 & 4.47 \\
  
  \bottomrule
  \end{tabular}%
  }
  
  \label{tab:performance_without_drift_H5}
\end{table}

\end{document}